\documentclass[a4paper,11pt]{article}
\usepackage{pos}
\usepackage{float}
\usepackage[caption=false]{subfig}
\usepackage{mwe}
\usepackage{multirow}
\usepackage{multicol}
\usepackage{comment}

\title{The $I=1/2$ and $3/2$ K-$\pi$ scattering length with domain wall fermions at physical pion mass with all-to-all propagators}
\ShortTitle{K$\pi$ scattering length at physical pion mass}

\author[a]{Nils Asmussen}
\author[b]{Felix Erben}
\author[a]{Jonathan Flynn}
\author[a,c]{Andreas J\"{u}ttner}
\author*[a]{Rajnandini Mukherjee}
\author[a]{Christopher T. Sachrajda}

\affiliation[a]{School of Physics and Astronomy, University of Southampton,
Southampton, UK}
\affiliation[b]{Higgs Centre for Theoretical Physics, University of Edinburgh, Edinburgh, UK}
\affiliation[c]{CERN, Theoretical Physics Department, Geneva, Switzerland}

\emailAdd{r.mukherjee@soton.ac.uk}

\abstract{We present our calculations for the $I=1/2$ and $3/2$ K-$\pi$ s-wave scattering length with physical quark masses, extracted from the interaction energy of Euclidean two-point functions. We use the domain wall fermion action with physical quark masses at a single lattice spacing. We are specifically interested in the systematic effects due to around-the-world terms on the overall determination of the scattering length. We present our progress and discuss the various systematic effects in our preliminary results.}

\FullConference{%
  The 39th International Symposium on Lattice Field Theory\\
  8-14 August, 2022\\
  Bonn, Germany
}

\begin{document}
\maketitle

\section{Introduction}
To further our understanding of strong interactions in the standard model, it is of interest to look at particle resonances at low energy, such as kaon-pion scattering. While this can be formulated using an effective-theory approach such as chiral perturbation theory \cite{ChiPT}, simulations of lattice QCD can provide both qualitative and quantitative insight from first principles. This is one of the cleanest scattering processes with regard to systematics and is a stepping stone towards high precision calculations that are particulary sensitive to certain systematic effects, such as in \cite{distillation} and \cite{distillation2}.

In this project, we compute the s-wave $K\pi$ scattering length in the two isospin channels: $I=1/2$ and $3/2$. There are previous determinations for these two quantities \cite{Sasaki, ETMC, Wilson} at unphysically heavy pion and/or kaon masses which use chiral perturbation theory to extrapolate the results to the physical pion and kaon mass. In our work, we calculate the scattering length directly at physical masses, using 23 configurations on a $48^3\times 96$ lattice with inverse spacing $a^{-1} = 1.730(4)$ GeV \cite{UKQCD}. We use the M\"{o}bius domain wall fermion action and an all-to-all setup with meson-field contractions for computing propagators. We compute a number of low modes of the Dirac operator exactly, and the high modes are stochastically sampled by inverting the Dirac operator on $Z_2\times Z_2$ noise sources. 
\section{Methodology}
We are interested in the s-wave scattering length, $a_0$, described by L\"{u}scher's formula \cite{Luscherformula} for interactions in a Euclidean volume of spatial extent $L$ as
\begin{align}
    \Delta E_{K\pi} &= -\frac{2\pi a_0}{\mu_{K\pi}L^3}\left(1 + c_1\frac{a_0}{L} +c_2\frac{a_0^2}{L^2}\right) + \mathcal{O}\Big(\frac1{L^6}\Big), \label{eqn:luscher}
\end{align}
where $c_1 = -2.837297$, $c_2 = 6.375183$ and $\mu_{K\pi}=m_\pi m_K/(m_{\pi}+m_K)$ is the reduced mass. Eqn. (\ref{eqn:luscher}) relates the scattering length to $\Delta E_{K\pi}$, the binding energy of the $K\pi$ composite state. This in turn can be extracted by studying the ground state of the 2-point function of some $K\pi$ operator. In our study, we choose the following $K\pi$ operator
\begin{align}
    O_{K\pi}(t) = K(t+\delta)\pi(t),
\end{align}
where the kaon and pion are separated in time  by $\delta=a$ in order to avoid unwanted contributions from additional Wick contractions of the correlated $Z_2$ noise sources. The corresponding 2-point function
\begin{align}
    C_{K\pi}(t) =& \langle O^{\text{snk}}_{K\pi}(t)O^{\dagger\text{src}}_{K\pi}(0)\rangle = \langle K(t+\delta)\pi(t)(K(\delta)\pi(0))^{\dagger}\rangle, \label{eqn:operators}
\end{align}
 has a straightforward spectral decomposition in infinite volume with ground state energy $E_{K\pi}=m_K+m_{\pi}+\Delta E_{K\pi}$. However, periodic boundary conditions introduce `around-the-world' effects, giving rise to additional terms 
\begin{align}
    C_{K\pi}(t) = &|\langle K \pi|K(\delta) \pi(0)| 0\rangle|^{2} \left(e^{-E_{K \pi} t} + e^{-E_{K \pi}(T-t)}\right) \label{term:cosh}\\ 
    &+|\langle \pi|K(\delta) \pi(0)| K\rangle|^2 e^{-m_{K} t} e^{-m_{\pi}(T-t)} \label{term:piKpiK}\\ 
    &+|\langle K|K(\delta) \pi(0)| \pi\rangle|^2 e^{-m_{\pi} t} e^{-m_{K}(T-t)} \label{term:KKpipi} \\
    &+\cdots \nonumber
\end{align}
where $T$ is the time extent of the lattice, and further terms are suppressed by multiples of pion and kaon masses. The 2-point function has a cosh-like term (\ref{term:cosh}) and two around-the-world (ATW) terms (\ref{term:piKpiK}) and (\ref{term:KKpipi}) corresponding to single meson states going around the periodic boundary. The binding energy, $\Delta E_{K\pi}$, appears in the cosh-like term and to extract it precisely from $C_{K\pi}$, we must account for the two ATW terms. Note that this is the case even with the modest statistical precision of our study.

Using all-to-all propagators should allow for low statistical error per configuration in $C_{K\pi}(t)$ by averaging multiple source planes. However, it is possible to have further noise cancellations on each configuration by instead studying the ratio
\begin{align}
    R_{K\pi}(t) = \frac{C_{K\pi}(t)}{C_K(t)C_{\pi}(t)} = \frac{\langle K(t+\delta)\pi(t)(K(\delta)\pi(0))^{\dagger}\rangle}{\langle\pi(t)\pi(0)^{\dagger}\rangle\langle K(t+\delta)K(\delta)^{\dagger}\rangle},
\end{align}
which also separates into a cosh-like piece, $R_{K\pi}^{\text{cosh}}(t)$, and an ATW piece $R_{K\pi}^{\text{ATW}}(t)$. Here we use the pion and kaon 2-point functions, which we separately fit to also extract the values of the pion and kaon masses and their amplitudes.

The ATW part of this ratio is given by
\begin{align}
    R^{\text{ATW}}_{K\pi}(t) = &\left|\frac{\langle \pi|K(\delta) \pi(0)| K\rangle}{\langle 0|K| K\rangle\langle\pi|\pi| 0\rangle}\right|^2 \frac{e^{-m_{K} t} e^{-m_{\pi}(T-t)}}{\left(e^{-m_{\pi}t}+e^{-m_{\pi}(T-t)}\right)\left(e^{-m_Kt}+e^{-m_K(T-t)}\right)} \\
    &+\left|\frac{\langle K|K(\delta) \pi(0)| \pi\rangle}{\langle 0|K| K\rangle\langle\pi|\pi| 0\rangle}\right|^2 \frac{e^{-m_{\pi} t} e^{-m_{K}(T-t)}}{\left(e^{-m_{\pi}t}+e^{-m_{\pi}(T-t)}\right)\left(e^{-m_Kt}+e^{-m_K(T-t)}\right)}.
\end{align}

All terms in this equation are fully determined by the information from the pion and kaon 2-point functions except for the matrix elements $\langle \pi|K(\delta) \pi(0)| K\rangle$ and $\langle K|K(\delta) \pi(0)| \pi\rangle$. In order to compute them, we separately construct correlation functions --- called $C_{\pi K\pi K}$ and $C_{KK\pi\pi}$ respectively --- that carry these matrix elements in their ground state contributions. In particular, we use the following 3-point functions
\begin{align}
    C_{\pi K\pi K} &= \langle \pi(\Delta)K(t+\delta)\pi(t)K(0)\rangle \label{eqn:piKpiK}\\
    &= \langle 0|\pi|\pi\rangle \langle \pi|K(\delta)\pi(0)|K\rangle \langle K|K|0\rangle  e^{-m_{\pi}(\Delta-t)}e^{-m_Kt}+\cdots,\nonumber\\
    C_{KK\pi\pi} &= \langle K(\Delta)K(t+\delta)\pi(t)\pi(0)\rangle \label{eqn:KKpipi}\\
    &= \langle 0|K|K\rangle \langle K|K(\delta)\pi(0)|\pi\rangle \langle\pi|\pi|0\rangle e^{-m_{\pi}t}e^{-m_K(\Delta-t)}+\cdots. \nonumber
\end{align}

The unknown matrix elements $\langle \pi|K(\delta) \pi(0)| K\rangle$ and $\langle K|K(\delta) \pi(0)| \pi\rangle$ can then be extracted by fitting these 3-point functions to their analytical forms. In this way we can now completely account for the ATW contributions and what remains is to extract the signal. As before, the signal lies in the cosh-like term
\begin{align}
    R_{K\pi}^{\text{cosh}}(t) = A\frac{\left(e^{-({m_{\pi}+m_K}+\Delta E_{K\pi})t} + e^{-(m_{\pi}+m_K+\Delta E_{K\pi})(T-t)}\right)}{\left(e^{-m_{\pi}t} + e^{-m_{\pi}(T-t)}\right)\left(e^{-m_Kt} + e^{-m_K(T-t)}\right)}, \label{eqn:cosh}
\end{align}
where all the terms are known except the amplitude, $A$, and $\Delta E_{K\pi}$ which we fit for. Using the fit value of $\Delta E_{K\pi}$ in L\"{u}scher's formula (\ref{eqn:luscher}), we are able to determine the value of the scattering length. The lattice quantities required for this analysis are the pion and kaon 2-point functions, the $K\pi$ 2-point function, and the two 3-point functions for determining the unknown ATW matrix elements.

\begin{figure}
\subfloat[D\label{sfig:D}]{\includegraphics[width=0.34\textwidth]{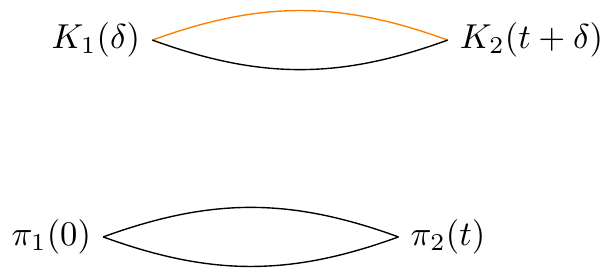}}
     \hfill
\subfloat[C\label{sfig:C}]{\includegraphics[width=0.29\textwidth]{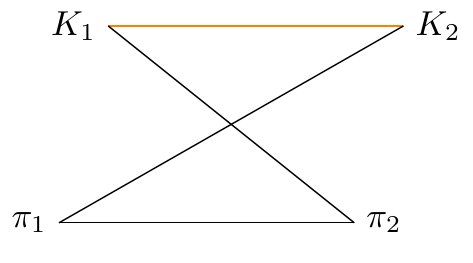}}
     \hfill
\subfloat[R\label{sfig:R}]{\includegraphics[width=0.29\textwidth]{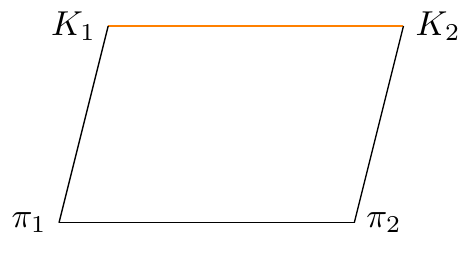}}
\caption{Diagrams corresponding to various Wick contractions that contribute to $C_{K\pi}$. Operators with subscripts $1$ and $2$ correspond to the source and sink operators respectively.}
\label{fig:DCR}
\end{figure}

Now that the procedure for calculating the scattering length is well defined, recall that we must do this for the two isospin channels $I=1/2$ and $3/2$. The above described procedure remains the same for the two isospin channels, however, in each case a given correlator is made up of weighted diagrammatic contributions corresponding to the isospin-dependent Wick contractions for that channel. The source operator in eqn. (\ref{eqn:operators}) is fixed to
\begin{equation}
    O_{K\pi}^{\text{src}}(t) = T\left[(\overline{u}\gamma_5 s)(t+\delta)(\overline{u}\gamma_5 d)(t)\right],
\end{equation}
while different sink operators are used to project onto the isospin $1/2$ or $3/2$ channels
\begin{align}
    O_{K\pi}^{\text{snk}\, I=1/2}(t) &= T\left[(\overline{d}\gamma_5 s)(t+\delta)\;(\overline{u}\gamma_5 d)(t) - \tfrac{1}{2}(\overline{u}\gamma_5 s)(t+\delta)\;(\overline{d}\gamma_5 d)(t)\right.\nonumber \\
    &\qquad+ \left.\tfrac{1}{2}(\overline{u}\gamma_5 s)(t+\delta)\;(\overline{u}\gamma_5 u)(t)\right], \\
    O_{K\pi}^{\text{snk}\, I=3/2}(t) &= T\left[(\overline{u}\gamma_5 s)(t+\delta)\;(\overline{u}\gamma_5 d)(t)\right].
\end{align}
The $C_{K\pi}$ correlation function is thus constructed from the combinations
\begin{align}
C_{K\pi}^{I=1/2} = D + \frac{1}{2}C - \frac{3}{2}R, \quad
C_{K\pi}^{I=3/2} = D - C,
\end{align}
where $D$, $C$, and $R$ are the direct, crossed and rectangular contractions, as illustrated in figure \ref{fig:DCR}. The same distinction is made when constructing the 3-point functions (\ref{eqn:piKpiK}) and (\ref{eqn:KKpipi}) for the two isospin channels.

\section{Results}
\begin{figure}[h]
    \subfloat{\includegraphics[width=0.49\textwidth]{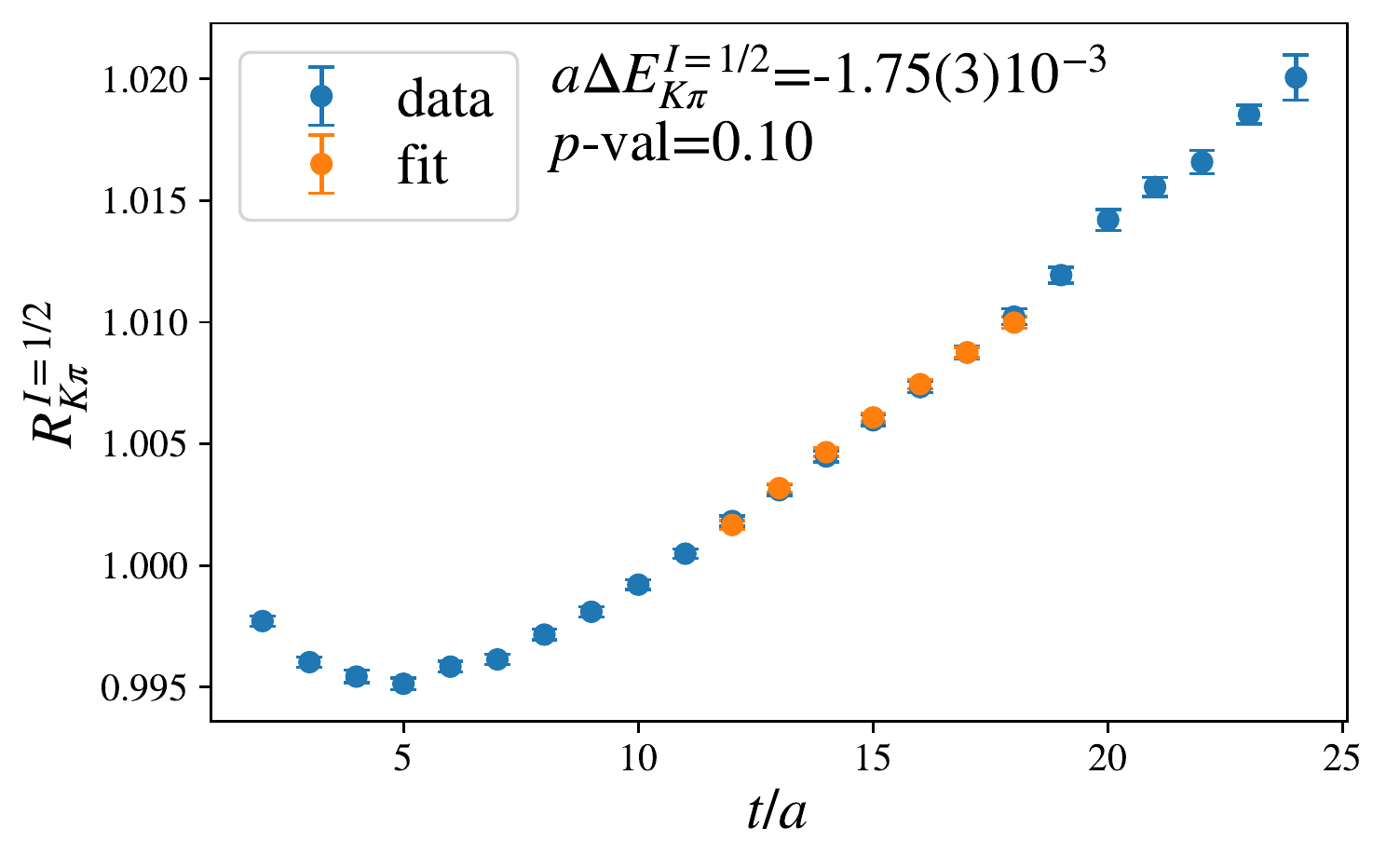}}
     \hfill
     \subfloat{\includegraphics[width=0.49\textwidth]{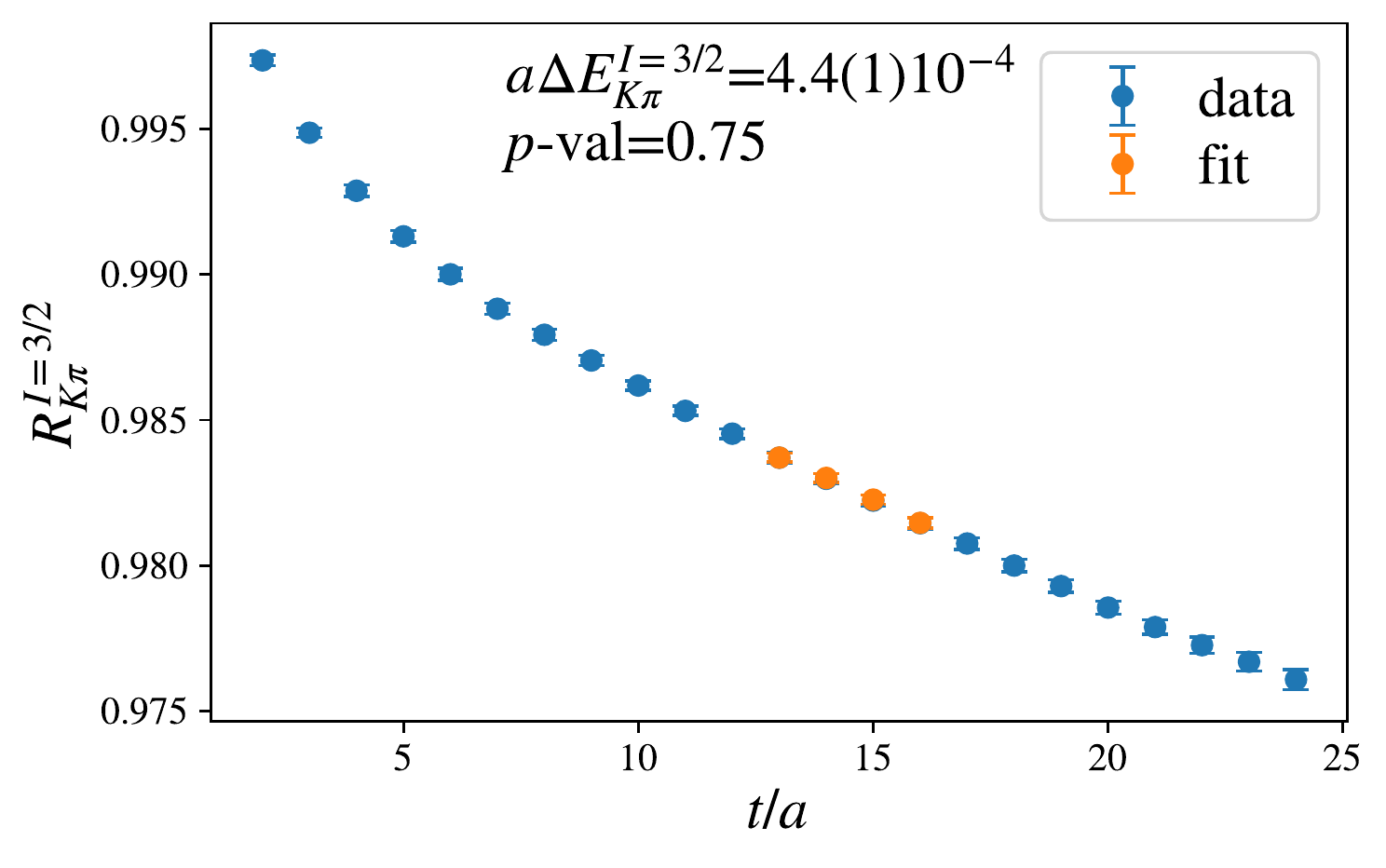}}
     \caption{Fits to $R_{K\pi}(t)$ data in the two isospin channels while accounting for ATW contributions (the orange points correspond to the fit interval). Fit values of $\Delta E_{K\pi}$ correspond to scattering lengths $m_{\pi}a_0^{I=1/2}=0.160(3)$ and $m_{\pi}a_0^{I=3/2}=-0.0377(2)$ using L\"{u}scher's formula in eqn. (\ref{eqn:luscher}).}
     \label{fig:fits}
\end{figure}

\begin{figure}[h!]
    \centering
    \subfloat{\includegraphics[width=0.48\textwidth]{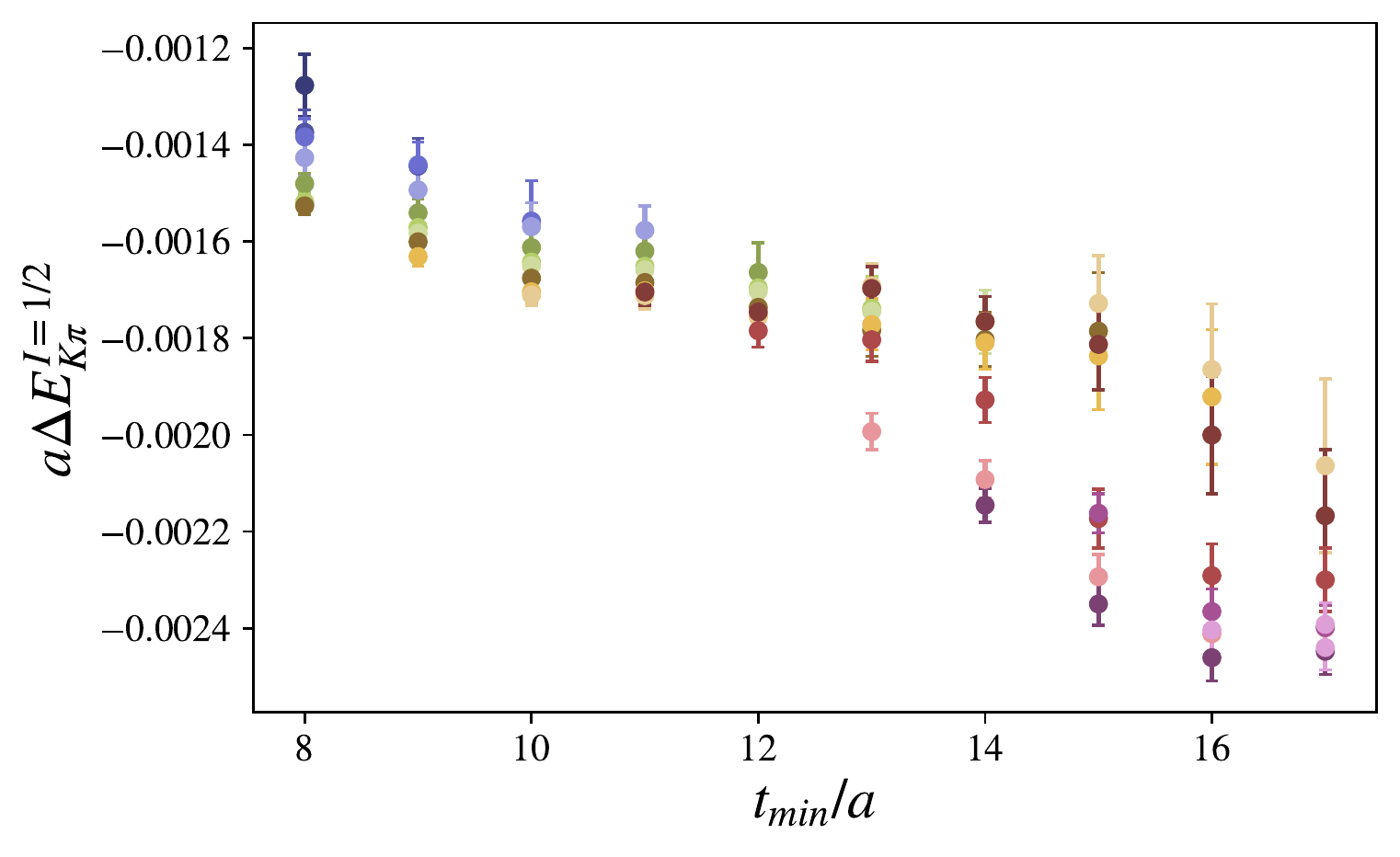}}
     \hfill
     \subfloat{\includegraphics[width=0.51\textwidth]{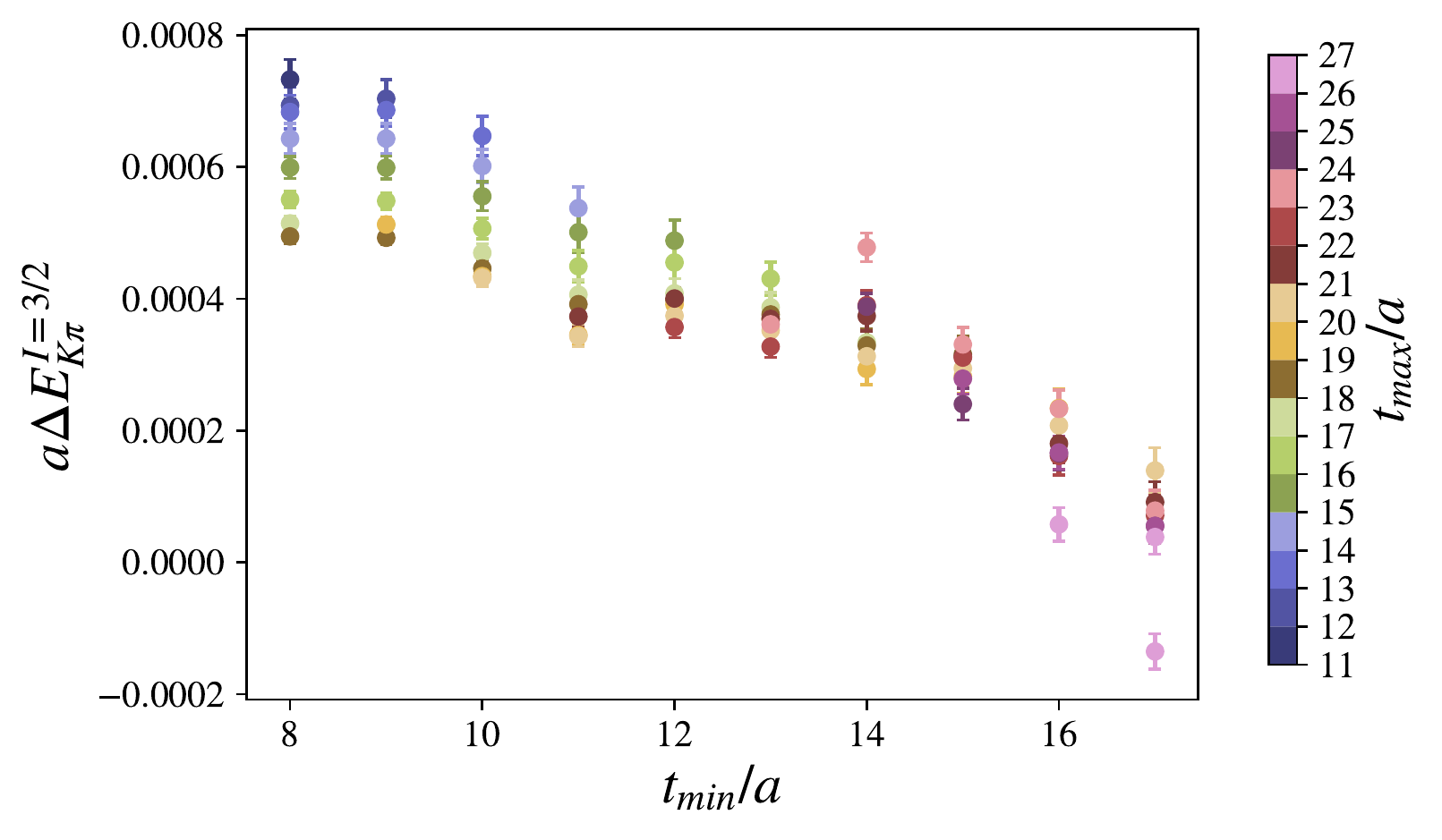}}
     \caption{Variation in fit value of $\Delta E_{K\pi}$ with fit intervals, extracted from $R_{K\pi}(t)$ for each isospin channel. This shows a strong systematic dependence of the results on the choice of both $t_{min}$ ($x$-axis) and $t_{max}$ (color in legend).}
     \label{fig:variations}
\end{figure}

Previous results for the $K\pi$ scattering length in the two isospin channels are listed in table \ref{tab:results}. We find sensible fits to $R_{K\pi}(t)$ in both channels, as shown in figure \ref{fig:fits}. However, we find a systematic drift in the fit results upon varying the $t_{min}$ and $t_{max}$ of the fit interval, shown in figure \ref{fig:variations}. Thus the fit result is not stable. To explain this trend in the fit value of $\Delta E_{K\pi}$ we consider the following possible explanations: effect of ATW contributions, contamination by excited states and availability of statistics.

\begin{table}[h]
    \centering
\begin{tabular}{|c|c|c|c|}
    \cline{2-4}
    \multicolumn{1}{c|}{} & $m_{\pi}/$MeV & $m_{\pi}a_0^{I=1/2}$ & $m_{\pi}a_0^{I=3/2}$ \\ \hline
    \multirow{2}{9em}{Sasaki (2014) \cite{Sasaki}} & extrap & 0.142(14)(27) & -0.0469(24)(20)  \\
    & 166 & 0.158(36) & -0.108(12)\\ \hline
    ETMC (2018) \cite{ETMC} & extrap$^{I=I/2}$/$\chi_{PT}^{I=3/2}$ & 0.163(3) & -0.059(2)\\ \hline   
    \multirow{2}{9em}{hadspec (2019) \cite{Wilson}} & 239 & 0.46(3) &   \\
    & 284 & 0.79(13) & \\ \hline
\end{tabular}
    \caption{Previous computations of $K\pi$ scattering length. The first error is statistical and the second is systematic error due to chiral extrapolation.}
    \label{tab:results}
\end{table}

\subsection{`Around-the-world' effects}

\begin{figure}[t]
    \centering
    \subfloat{\includegraphics[width=0.49\textwidth]{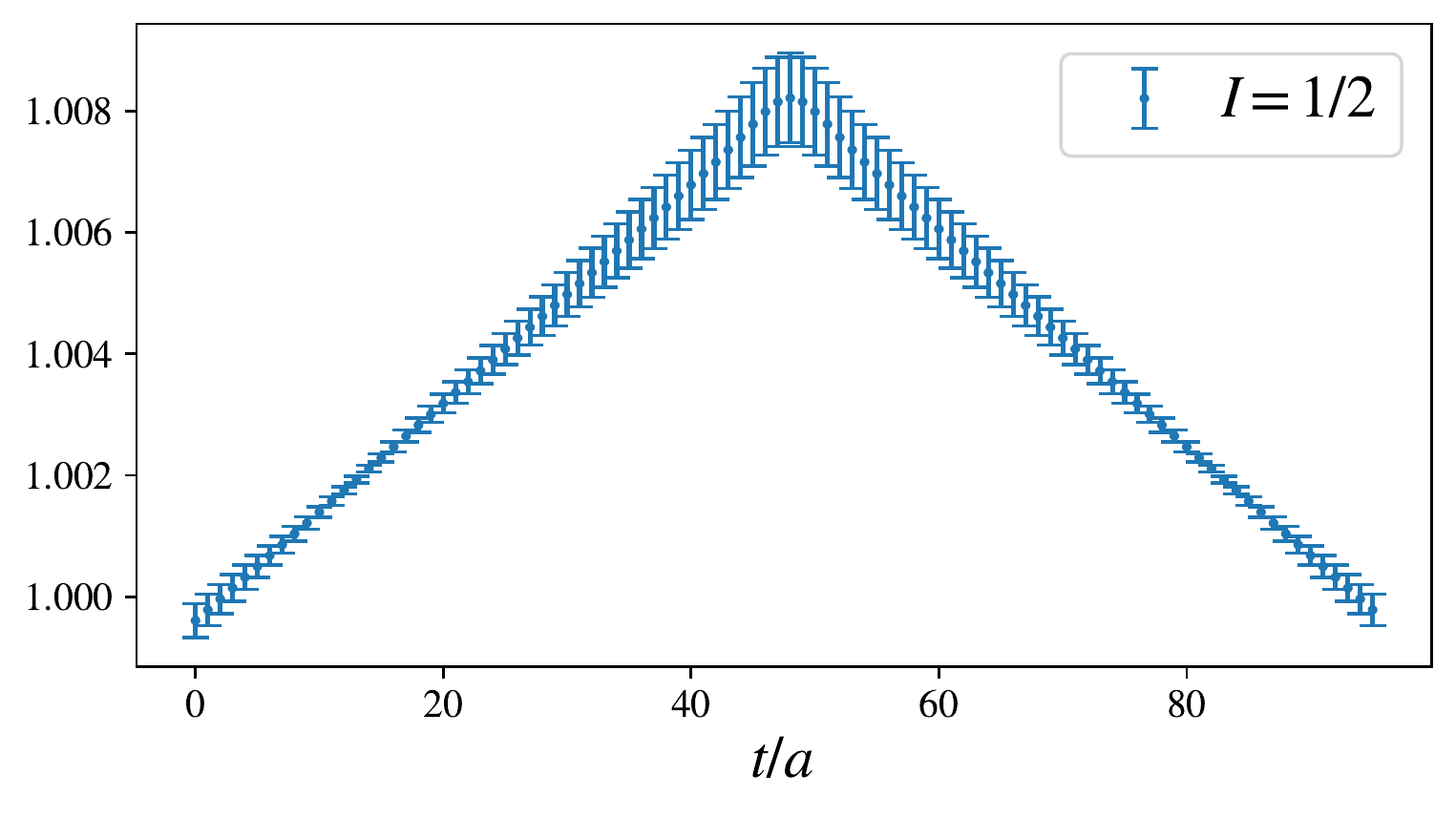}}
     \hfill
     \subfloat{\includegraphics[width=0.49\textwidth]{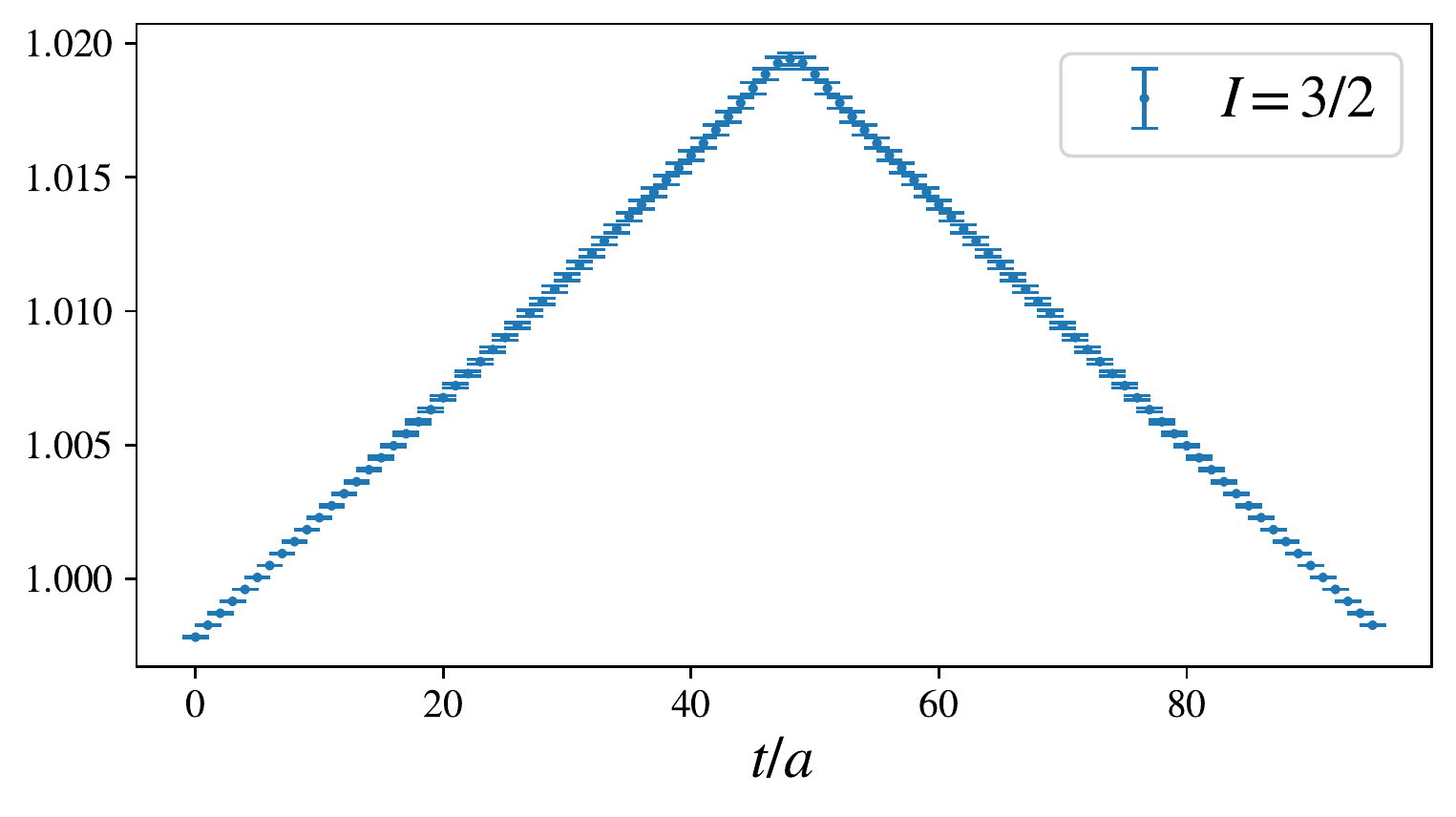}}
     \caption{Effect of the ATW terms in fitting $R_{K\pi}(t)$. These plots show the time-dependence of the ratio of the cosh-like component of fits using two different ansatz: $R_{K\pi}(t) = R_{K\pi}^{cosh}(t)$ and $R_{K\pi}(t) = R_{K\pi}^{cosh}(t)+R_{K\pi}^{ATW}(t)$.}
     \label{fig:ratio}
\end{figure}
\begin{figure}[t]
    \centering
    \subfloat{\includegraphics[width=0.49\textwidth]{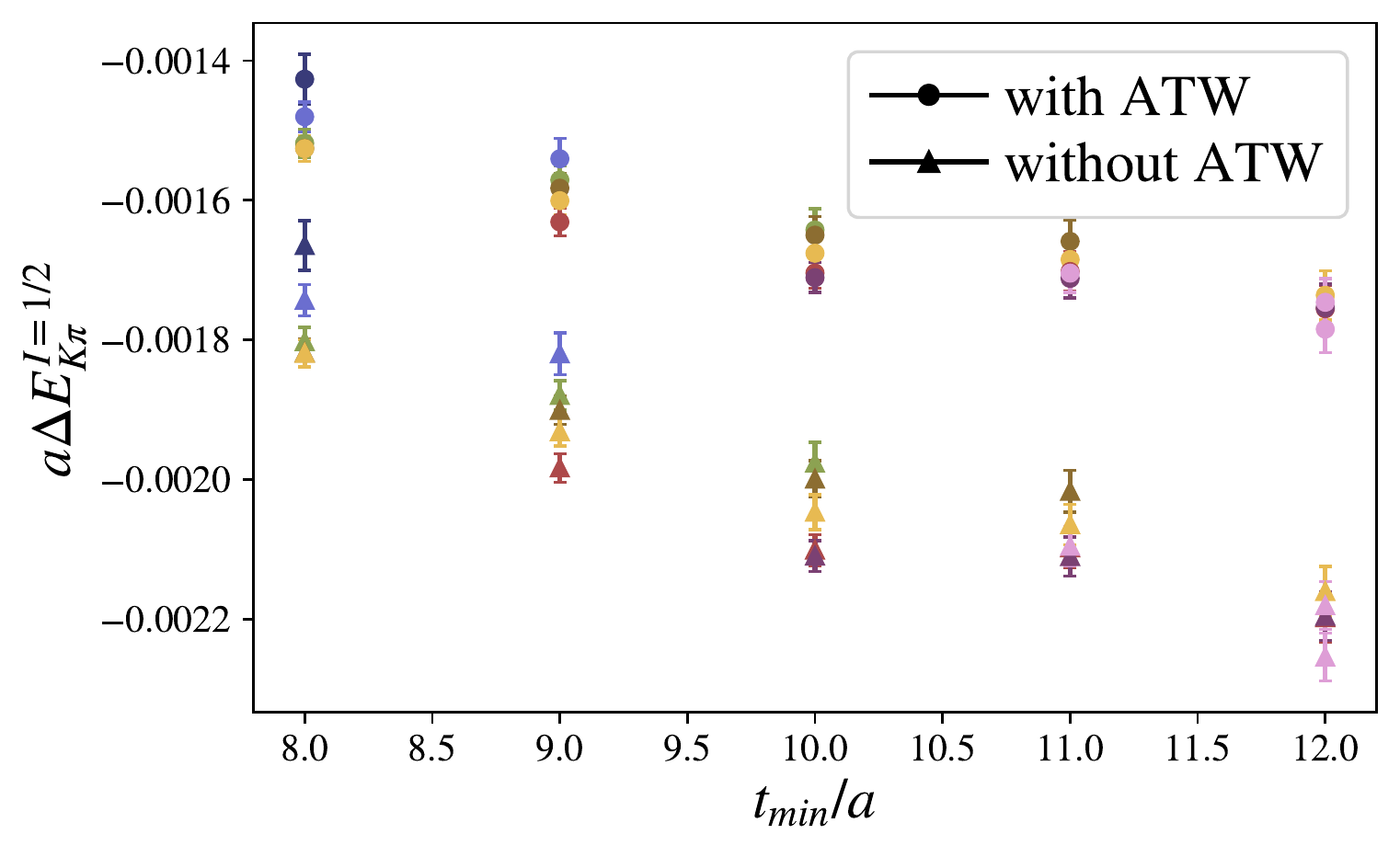}}
     \hfill
     \subfloat{\includegraphics[width=0.49\textwidth]{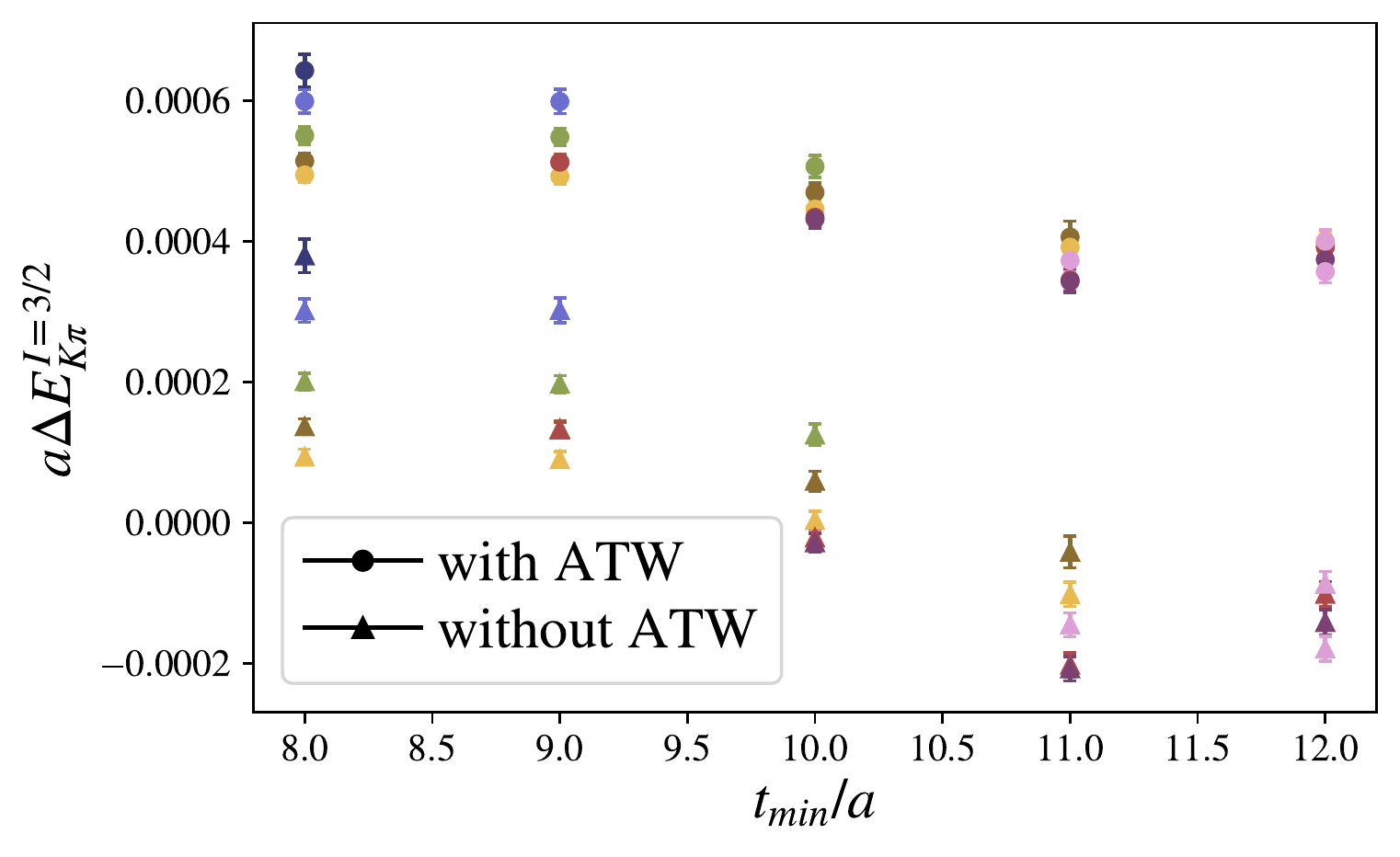}}
     \caption{Variation in fit value of $\Delta E_{K\pi}$ with fit intervals, extracted from $R_{K\pi}(t)$ from two ansatz, one accounting for ATW contributions, and the other not. It can be seen here that the ATW terms have a constant additive contribution to the value of $\Delta E_{K\pi}$.}
     \label{fig:ATW}
\end{figure}

By looking at the functional form of $R_{K\pi}^{\text{ATW}}(t)$, eqns. (\ref{eqn:piKpiK}) and (\ref{eqn:KKpipi}), it is evident that ATW terms have time-dependent contributions. As the systematic trend we see in the fits to $R_{K\pi}(t)$ is also time-dependent, it is worthwhile to study the effect of these ATW terms in the overall determination of $\Delta E_{K\pi}$. We investigate this effect by studying the cosh-like component of fits done with and without accounting for the ATW terms. This is to quantify how the ATW terms are absorbed into the cosh only ansatz when the ATW terms are unaccounted for, as shown in figure \ref{fig:ratio}. This tells us that the ATW terms have a time-dependent linear contribution at all times. To understand their effect on the value of $\Delta E_{K\pi}$, we look at eqns. (\ref{eqn:piKpiK}-\ref{eqn:cosh}) at small $t$, where
\begin{align}
    R_{K\pi}(t) \approx A + B\Delta E_{K\pi}t + C_{ATW}t + \mathcal{O}(t^2)
\end{align}
 which implies that the ATW terms shift the value of $\Delta E_{K\pi}$ when they are not accounted for. This shift is shown in figure \ref{fig:ATW} for both isospin channels. The effect of the ATW terms is particularly important in the $I=3/2$ channel where the shift is comparable in magnitude to $\Delta E_{K\pi}$.

ATW terms are substantial and present at all times, which is a particular feature of our study owing to the (low) physical pion mass which allows the ATW terms to be more dominant. It is therefore important to account for these terms in any further studies made at the physical point for a reliable determination of $\Delta E_{K\pi}$. However, as shown in figure \ref{fig:ATW}, they lead to a constant shift and do not account for the time-dependence.

\subsection{Excited states}
\begin{figure}[t]
    \centering
    \subfloat{\includegraphics[width=0.49\textwidth]{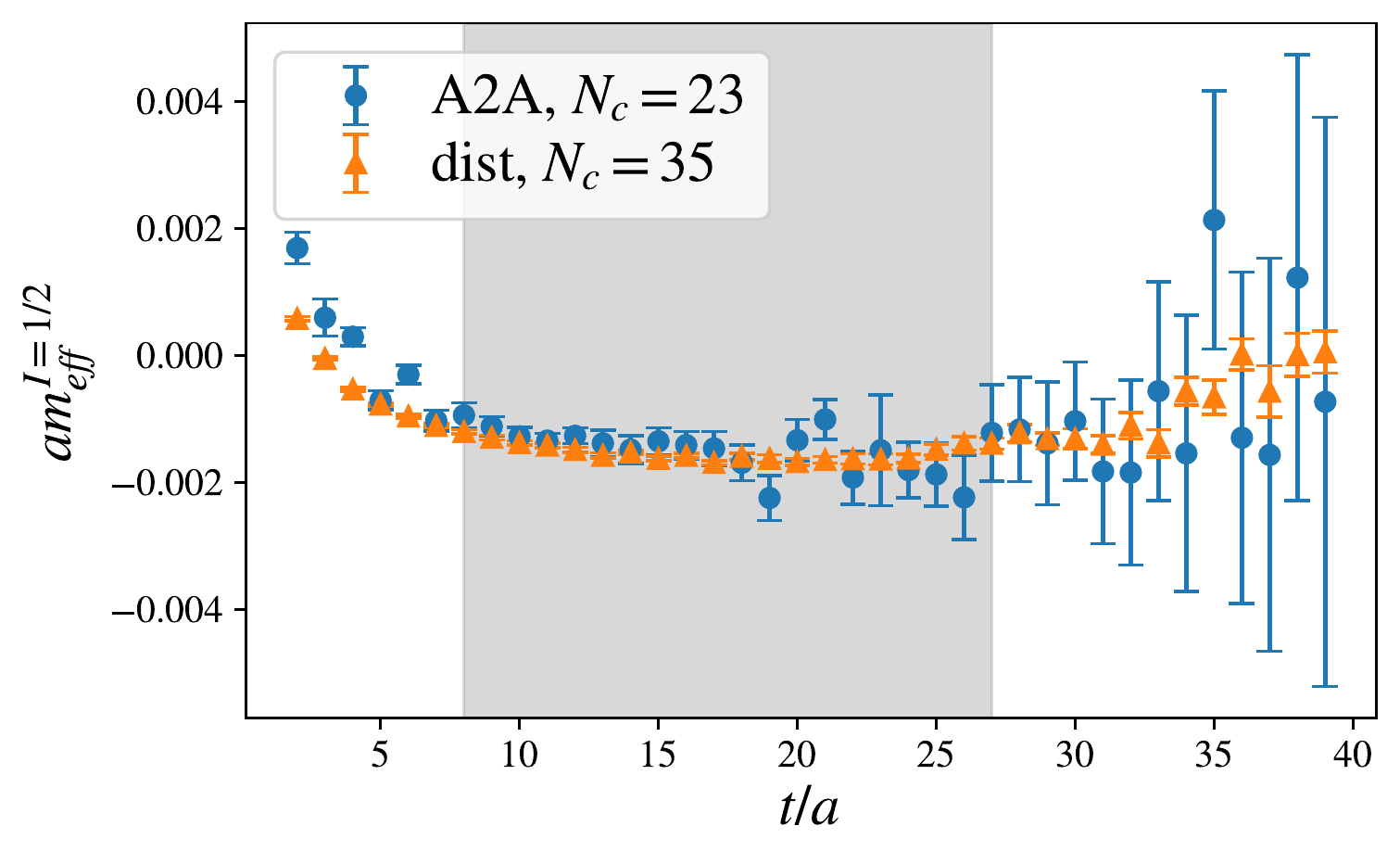}}
     \hfill
     \subfloat{\includegraphics[width=0.49\textwidth]{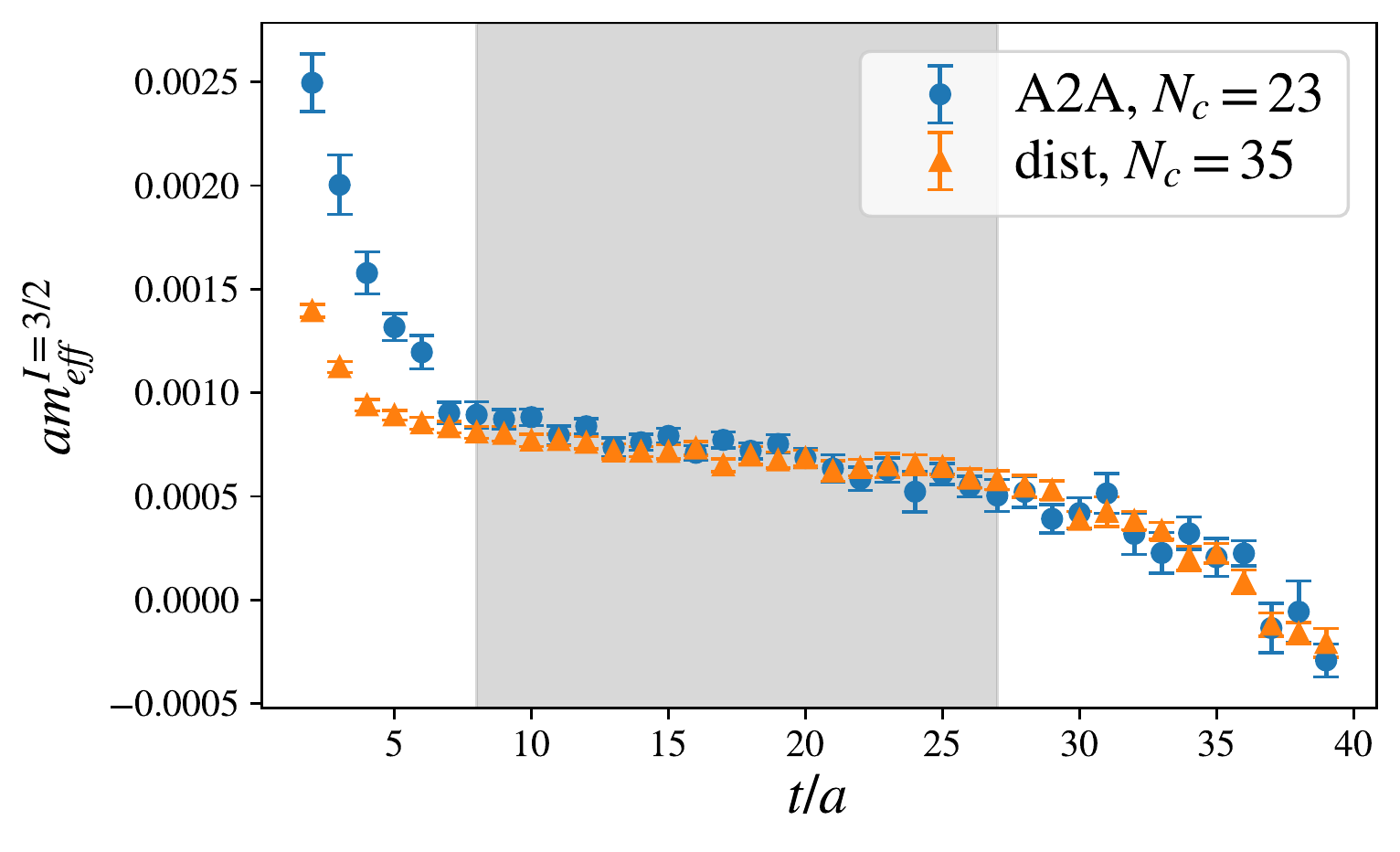}}
     \caption{Effective mass plots for $R_{K\pi}(t)$ in the two isospin channels, comparing the contamination of excited states in the all-to-all data compared to data generated via distillation. The grey region corresponds to the region where fits are made. $N_c$ refers to the number of configurations used.}
     \label{fig:ES}
\end{figure}

To investigate further the source of the time-dependent trend in the fit values of $\Delta E_{K\pi}$, we next study the effect of excited states in our $R_{K\pi}$ data. We do this by looking at the effective mass plot for $R_{K\pi}$ using our data, generated using all-to-all propagators, and data generated using a distillation procedure \cite{distillation} where the ground state of a $K\pi$ operator is determined to a greater accuracy. This is shown in figure \ref{fig:ES}, which reveals excited state contamination at low times. However these effects die out at larger times, and are negligible in the range where we perform fits (the grey area in the plots). While the distillation data has greater precision in comparison to the all-to-all data, this comparison shows us that the systematic trend we see in the fit value of $\Delta E_{K\pi}$ is unlikely a consequence of excited state effects.

\subsection{Statistics}
$\Delta E_{K\pi}$ is an extremely delicate quantity in comparison to the other observables we compute in this analysis. For example, the pion and kaon masses on the lattice are $am_{\pi}=0.0803(2)$ and $am_K=0.2887(2)$, and therefore $a\Delta E_{K\pi}$ is of the order of magnitude of the error in these masses. In our study we attempt to extract this using 23 configurations. While the all-to-all setup allows us to compute propagators with relatively low statistical error due to availability of a large number of source planes per configuration, the overall number of configurations is a limiting factor. Figure \ref{fig:ES} also indicates this effect of statistics as the distillation study has more configurations ($N_c=35$). However, comparison of the effective masses suggests it to be unlikely that the time-dependent trend in the fit for $\Delta E_{K\pi}$ is a direct effect of statistics, especially in the $I=3/2$ channel where the all-to-all data is in good agreement with the distillation data. We are therefore in need of larger statistics to pin down the systematics effects we see.

\section{Conclusion}
In this study we have attempted to calculate the $K\pi$ scattering length at physical pion mass. As this is one of the first calculations done directly at the physical point for both isospin channels, we encounter new features and new issues. The most interesting feature of our study is the effect of around-the-world contributions, which are substantial due to the light pion. This effect is significant at all times on the lattice especially when extracting a small and sensitive quantity such as the binding energy, $\Delta E_{K\pi}$. The time-dependent trend in the value of this quantity remains unexplained and may be a combination of effects considered above or an indicator of physics that has not been taken into account. We look forward with interest towards other analyses carried out at the physical point, where we expect around-the-world effects to play an important role in the study of scattering lengths.

\section{Acknowledgements}
The distillation data is shared by Nelson Lachini, Antonin Portelli, and F.E. This work used the DiRAC Extreme Scaling service at the University of Edinburgh, operated by the Edinburgh Parallel Computing Centre on behalf of the STFC DiRAC HPC Facility (www.dirac.ac.uk) under project codes dp008 and dp207 using Grid and Hadrons. This equipment was funded by BEIS capital funding via STFC capital grant ST/R00238X/1 and STFC DiRAC Operations grant ST/R001006/1. DiRAC is part of the National e-Infrastructure. R.M. is supported by the Presidential Scholarship from the University of Southampton. C.T.S. is partially supported by an Emeritus Fellowship from the Leverhulme Trust and by STFC (UK) grants ST/P000711/1 and ST/T000775/1. A.J. and J.M.F. also acknowledge funding by  ST/P000711/1 2017-2021 and ST/T000775/1. F.E. is supported in part by UK STFC grant ST/P000630/1. F.E. also received funding from the European Research Council (ERC) under the European Union’s Horizon 2020131 research and innovation programme under grant agreement No 757646.


\begin{thebibliography}{99}

\bibitem{ChiPT}
M. Jamin \textit{et al.}, \emph{S-wave Kpi scattering in chiral perturbation theory with resonances,  \href{https://arxiv.org/abs/hep-ph/0006045}{Nucl. Phys. B \normalfont{\textbf{587}}, 331-362 (2000)}}

\bibitem{distillation}
N. Lachini \textit{et al.}, \emph{$K\pi$ scattering at physical pion mass using distillation,  \href{https://inspirehep.net/literature/1993866}{PoS LATTICE2021 \normalfont{\textbf{435}} (2022)}}

\bibitem{distillation2}
F. Erben \textit{et al.}, \emph{Exploring distillation at the SU(3) flavour symmetric point,  \href{https://arxiv.org/abs/2211.15627}{PoS LATTICE2022}}

\bibitem{Sasaki}
K. Sasaki \textit{et al.}, \emph{Scattering lengths for two pseudoscalar meson systems, \href{https://journals.aps.org/prd/abstract/10.1103/PhysRevD.89.054502}{Phys. Rev. D \normalfont{\textbf{89}}, 054502 (2014)}}

\bibitem{ETMC}
C. Helmes \textit{et al.}, \emph{Hadron-Hadron Interactions from $N_f=2+1+1$ Lattice QCD: $I=3/2$  $\pi K$ Scattering Length, \href{https://journals.aps.org/prd/abstract/10.1103/PhysRevD.98.114511}{Phys. Rev. D \normalfont{\textbf{98}}, 114511 (2018)}}

\bibitem{Wilson}
D. Wilson \textit{et al.}, \emph{The quark-mass dependence of elastic $\pi K$ scattering from QCD,  \href{https://journals.aps.org/prl/abstract/10.1103/PhysRevLett.123.042002}{Phys. Rev. Lett. \normalfont{\textbf{123}}, 042002 (2019)}}

\bibitem{UKQCD}
T. Blum \textit{et al.}, \emph{Domain wall QCD with physical quark masses,  \href{https://journals.aps.org/prd/abstract/10.1103/PhysRevD.93.074505}{Phys. Rev. D \normalfont{\textbf{93}}, 074505 (2016)}}


\bibitem{Luscherformula}
M L\"{u}scher, \emph{Volume dependence of the energy spectrum in massive quantum field theories, \href{https://link.springer.com/article/10.1007/BF01211097}{Commun. Math. Phys. \normalfont{\textbf{104}} (1986) 177.}}

\end{thebibliography}
\end{document}